\documentclass{article}

\usepackage{arxiv}

\usepackage[utf8]{inputenc} 
\usepackage[T1]{fontenc}    
\usepackage{hyperref}       
\usepackage{url}            
\usepackage{booktabs}       
\usepackage{amsfonts}       
\usepackage{nicefrac}       
\usepackage{microtype}      
\usepackage{calc}
\usepackage{epsfig}
\usepackage{subcaption}
\usepackage{calc}
\usepackage{amssymb}
\usepackage{amstext}
\usepackage{amsmath}
\usepackage{amsthm}
\usepackage{multicol}
\usepackage{pslatex}
\usepackage{apalike,pifont}
\usepackage{graphicx}
\usepackage{doi}
\usepackage{cite}
\usepackage{amsfonts}
\usepackage{algorithmic}
\usepackage{graphicx}
\usepackage{textcomp}
\usepackage{multirow}
\usepackage{xcolor}
\usepackage{booktabs}
\usepackage{hyperref}
\usepackage{lscape} 
\usepackage{threeparttable}

\title{To See or Not to See: A Privacy Threat Model for Digital Forensics in Crime Investigation}

\author{ \href{https://orcid.org/0000-0002-7045-0213}{\includegraphics[scale=0.06]{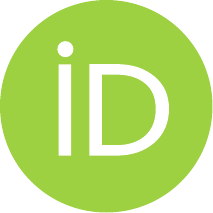}\hspace{1mm}Mario Raciti} \\
	IMT School for Advanced Studies Lucca\\
	Lucca, Italy \\
	\texttt{mario.raciti@imtlucca.it} \\
	\And
	\href{https://orcid.org/0009-0005-7888-2191}{\includegraphics[scale=0.06]{orcid.pdf}\hspace{1mm}Simone Di Mauro} \\
	Università di Catania\\
	Catania, Italy \\
	\texttt{uni360952@studium.unict.it} \\
    \And
    \href{https://orcid.org/0000-0001-6597-2271}{\includegraphics[scale=0.06]{orcid.pdf}\hspace{1mm}Dimitri Van Landuyt} \\
	KU Leuven\\
	Leuven, Belgium \\
	\texttt{dimitri.vanlanduyt@kuleuven.be} \\
    \And
	\href{https://orcid.org/0000-0002-7615-8643}{\includegraphics[scale=0.06]{orcid.pdf}\hspace{1mm}Giampaolo Bella} \\
	Università di Catania\\
	Catania, Italy \\
	\texttt{giampaolo.bella@unict.it} \\
}

\renewcommand{\shorttitle}{To See or Not to See: A Privacy Threat Model for DFCI}

\begin{document}
\maketitle

\begin{abstract}
Digital forensics is a cornerstone of modern crime investigations, yet it raises significant privacy concerns due to the collection, processing, and storage of digital evidence. Despite that, privacy threats in digital forensics crime investigations often remain underexplored, thereby leading to potential gaps in forensic practices and regulatory compliance, which may then escalate into harming the freedoms of natural persons.
With this clear motivation, the present paper applies the SPADA methodology for threat modelling with the goal of incorporating privacy-oriented threat modelling in digital forensics. 
As a result, we identify a total of 298 privacy threats that may affect digital forensics processes through crime investigations. Furthermore, we demonstrate an unexplored feature on how SPADA assists in handling domain-dependency during threat elicitation. This yields a second list of privacy threats that are universally applicable to any domain. We then present a comprehensive and systematic privacy threat model for digital forensics in crime investigation.
Moreover, we discuss some of the challenges about validating privacy threats in this domain, particularly given the variability of legal frameworks across jurisdictions. 
We ultimately propose our privacy threat model as a tool for ensuring ethical and legally compliant investigative practices.
\end{abstract}

\keywords{threat modeling, privacy, investigation, SPADA}

\section{Introduction}
\label{sec:introduction}

Digital forensics play an increasingly important role in modern crime investigations, yet simultaneously also introduce significant privacy concerns. Investigators must access, process, and store vast amounts of digital evidence, raising risks of misuse and unauthorised data processing.
Recent incidents highlight the real-world implications of privacy breaches in digital investigations. For instance, in Serbia, authorities have reportedly used advanced mobile forensics tools alongside spyware to surveil journalists and activists, extracting data without consent and raising significant ethical and legal concerns~\cite{guardian}.
Furthermore, discussions about governmental demands for backdoor access to encrypted user data --for example in messaging applications-- have intensified. Such measures, intended for national security purposes, pose risks to individual privacy and data security, leading to debates about the extent of governmental surveillance powers~\cite{apnews}.
Existing regulations such as the General Data Protection Regulation (GDPR) and law enforcement directives such as the Police Directive (EU 2016/680), provide foundational legal safeguards, yet they are often challenged by evolving forensic techniques and the complexities of cross-border investigations.

This paper presents a comprehensive privacy threat model for digital forensics in crime investigation to support forensic investigators in mitigating privacy risks while preserving the evidentiary integrity of forensic processes, and raising awareness among legal professionals and defendants regarding potential privacy violations within forensic investigations.

We apply a state-of-the-art privacy threat modelling methodology called SPADA~\cite{spada}, which is particularly suited for creating domain-specific threat models by combining domain-independent and domain-specific knowledge resources.

The rest of this paper is structured as follows. Section~\ref{sec:related-work} outlines the related work. Section~\ref{sec:spada} briefly recalls the key elements of the SPADA methodology and Section~\ref{sec:application} applies SPADA to the domain of digital forensics in crime investigation. Section~\ref{sec:discussion} discusses the results and ~\ref{sec:conclusions} concludes.

\subsection{Context and Motivation}
\label{subsec:motivation}

Privacy threats in digital forensics may arise from various sources, including unauthorised access to case-related data, data retention beyond legally permissible limits, improper access control mechanisms, and procedural oversights. These threats can compromise the rights of individuals involved in an investigation, including suspects, witnesses, and victims.
Additionally, unaddressed legal issues (among which related to privacy and data protection) may inadvertently harm the evidentiary value of collected digital forensics material. 

Moreover, our previous work~\cite{raciti2024} discussed a visual formalisation of digital forensics in crime investigation (DFCI), based on the claim that the overall process, from the news of a committed crime to a sentence in a court, can be considered a protocol. 
The process  must adhere to specific rules to ensure the fundamental rights of individuals (suspects) and ultimately protect society against systemic abuse. 
Briefly, our previous work delineated and formalised the protocols that compose digital forensics in crime investigation as Message Sequence Charts (MSCs). 
While this formalisation identified functional requirements, it left non-functional aspects, such as privacy and compliance risks, for subsequent investigation. 
Therefore, a logical extension of our previous work involves systematically eliciting non-functional requirements to complement the formalisation of DFCI. 
One viable approach to achieving this is through a structured threat modelling, where a comprehensive list of threats serves as the foundation for deriving privacy requirements in digital forensics crime investigations.

Threat modelling provides a systematic approach to identifying, categorising, and mitigating threats. The motivation for this paper also rests on the fact that a comprehensive privacy threat model for DFCI will ensure forensic processes are both effective and privacy-preserving, and appropriate trade-offs between these sometimes competing concerns can be found.

\subsection{Research Question and Contributions}
\label{sec:contributions}

Following the context and motivations given above, this paper focuses on and addresses the following research question:

\begin{quote}
\textbf{RQ}: What are the privacy threats in a digital forensics crime investigation?
\end{quote}

To answer this question, we apply the SPADA methodology for threat modelling~\cite{spada} in the digital forensics domain.
The main contributions of this paper can be summarised as follows.
We elicit a list of privacy threats in the domain of digital forensics in crime investigation. 
While we cannot claim that such list is complete, it extends existing perspectives on privacy concerns. In addition, we demonstrate an unexplored feature on how SPADA assists in handling domain-dependency during threat elicitation.
As a result, we present a comprehensive and systematic privacy threat model for DFCI.

By integrating privacy considerations into structured threat modelling for digital forensics in crime investigation, this paper aims to promote forensic investigations while ensuring compliance with privacy safeguards and ethical standards, thus safeguarding the privacy of individuals involved. The results are available online~\cite{repo}.

\section{Related Work}
\label{sec:related-work}

The intersection between digital forensics and threat modelling has been studied in several aspects.

Iqbal et al.~\cite{iqbal2020} established the importance of forensic evidence and how it can aid threat modelling by providing more comprehensive threat intelligence. Tok et al.~\cite{tok2023} applied the STRIDE framework and the Microsoft Threat Modelling Tool to help investigators understand what crimes could have been committed and what evidence would be required in their investigation work.
However, (privacy) threat modelling the digital forensics process remains an underexplored topic.

Chaure and Mane~\cite{Chaure_Mane_2023} acknowledged the significant challenges of preventing privacy breaches during a digital forensic investigation. The authors proposed an intelligent framework for preserving data privacy during forensic investigations, incorporating machine learning for identifying evidence while protecting non-evidential private files. Yet, the work emphasises that shielding data privacy without compromising the investigation's efficiency remains an open problem. Furthermore, the authors identify the process to obtain consent from individuals in possession of private information as being laborious and time-consuming.

Schaik et al.~\cite{Schaik} focused on how privacy rights and the use of Privacy-Enhancing Technologies (PETs) influence police digital forensics practices. Also in this case, the authors discussed the necessity for obtaining consent before searching a victim's digital device, as outlined by the Digital Processing Notice (DPN) in the UK. 

In the work by Seyyar and Geradts~\cite{Seyyar}, the focus shifted toward Privacy Impact Assessments (PIAs), which are mandatory under the GDPR. The authors addressed the increasing collection of personal data from seized digital devices, examining privacy risks in large-scale digital forensic investigations. While this work provides a methodology for conducting PIAs, it does not offer a structured approach to threat modelling. In fact, the authors concluded that a more structured threat modelling methodology is needed to consistently address privacy concerns in forensic investigations.

Rowe~\cite{rowe} focused on a broad overview of the privacy challenges in digital forensics. The paper discussed the potential for surreptitious searches and the sale of forensic data, emphasising the ineffectiveness of current privacy laws like GDPR in mitigating these risks. The concept of forensic transparency and the need for forensic privacy policies are suggested as ways to improve user control over their data, but these suggestions focus more on policy than systematic threat analysis.

To the best of our knowledge, no existing work has systematically and comprehensively modeled privacy threats in digital forensics crime investigations.

As we adopt SPADA as the main methodology for constructing this artifact, we first discuss the methodology in Section~\ref{sec:spada}. 
Then, we discuss its application to DFCI in Section~\ref{sec:application}.

\section{A Primer on SPADA}
\label{sec:spada}

This section outlines the main features of the SPADA methodology for threat modelling~\cite{spada}.
SPADA incorporates both domain-independent and domain-specific knowledge, ensuring a structured approach to indentifying threats. 
SPADA adopts five key variable elements that contribute to model security and privacy threats, and names them variables. 
These variables together constitute the SPADA acronym:

\begin{itemize}
    \item \textbf{S}ource of documentation --- The source(s) from which threats and assets may be derived, i.e., internal, external, or hybrid.
    Sources are, for example, public standards as well as institutional policies and semi-structured information gathered through interviews.
    \item \textbf{P}roperty --- The specific version of the target property that the threat modelling effort defines. 
    Properties are, for example, soft privacy, hard privacy~\cite{danezisg, hoepman} and cybersecurity.
    \item \textbf{A}pplication domain --- The domain-dependency that the threat modelling efforts defines.
    Domain-dependency refers, for example, to smart cars or smart homes (hence threats are instantiated on domain assets), else to no domain, or domain-independency (hence threats are universally applicable).
    \item \textbf{D}etail (level of) --- The detail refers to the style of each statement describing a threat or asset. 
    The detail is, for example, abstract (when threats are specified in broad, generalised terms) or detailed (when otherwise).
    \item \textbf{A}gent(s) raising the threats --- Agents here refer to the active entities that provide root cause of the threats. 
    Such agents are, for example, the attacker, the data controller, and the data processor, 
    or some combination of them, such as the controller and the processor.
\end{itemize}

Furthermore, the SPADA methodology consists of the following four operations:

\begin{enumerate}
    \item \textit{Combine}: it instantiates a domain-independent threat with the domain-specific assets that may be affected by such threat, typically by continuing the threat description with an explicit reference to the asset.
    \item \textit{Embrace}: it merges multiple threats into a single threat, typically by a new threat description that embodies the semantics of the given threats.
    \item \textit{Rename}: it modifies the description of a threat. This results particularly useful when the analyst wants to further refine a threat description.
    \item \textit{Discard}: it excludes a threat from the current analysis. The discarded threat may be kept in a reserve list for potential future review.
\end{enumerate}

In summary, the SPADA methodology follows these steps:

\begin{enumerate}
\setlength{\itemindent}{8mm}
    \item[\textbf{Step 0}] \textbf{Variable Setup:}~Defines the values for the five variables.
    \item[\textbf{Step 1}] \textbf{Domain-Independent Threat Elicitation:}~Extract threats that are universally applicable from relevant document sources.
    \item[\textbf{Step 2}] \textbf{Domain-Dependent Asset Collection:}~Identifies assets relevant to the specific application domain.
    \item[\textbf{Step 3}] \textbf{Domain-Dependent Threat Elicitation:}~Combines domain-independent threats with domain-specific assets, thereby instantiating threats into domain-dependent threats.
\end{enumerate}

SPADA can generate a list of domain-independent threats at Step 1. 
If domain specificity is required, it proceeds to Steps 2 and 3, ultimately producing domain-dependent threats by contextualising them within the identified assets.
This capability makes SPADA particularly suited for the purpose of creating a threat model specific to Digital Forensics in Crime Investigation (DFCI).

\section{Application of SPADA in DFCI}
\label{sec:application}

This section applies the SPADA methodology to the domain of digital forensics in crime investigation using a two-fold approach. 
The results are available online~\cite{repo}.

\subsection{Step 0:~Variable Setup}

\begin{table}[ht]
    \centering
    \caption{SPADA Variable setup for DFCI.}
    \label{fig:variables}
\begin{threeparttable}

\begin{tabular}{ p{0.05\linewidth}p{0.85\linewidth}}\toprule
    \textbf{S} & \emph{External}: Seyyar, Chaure, Rowe, Shaik, ISO, CoE DF, CoE EEG, IPOL, NIST, NIJ\\
    \textbf{P}& \emph{Soft and Hard Privacy}\\
    \textbf{A} & \emph{Domain-Dependent}: Digital Forensics in Crime Investigation \\
    \textbf{D} & \emph{Abstract}\\
    \textbf{A} & \emph{Attacker, Data Controller/Processor, Third Party}\\\bottomrule
\end{tabular}
\footnotesize
\begin{tablenotes}[]
        \item[ISO] ISO/IEC 27037:2012 (Guidelines for identification, collection, acquisition and preservation of digital evidence)~\cite{iso27037}
        \item[CoE DF] Council of Europe Electronic Evidence Guide (Version 3.0)~\cite{coe_eeg}.
        \item[CoE EEG] Council of Europe Electronic Evidence Guide (Version 3.0)~\cite{coe_eeg}
        \item[IPOL] EU Directorate General for Internal Policies of the Union Criminal Procedural Laws Across the European Union~\cite{ipol}. 
        \item[`NIJ]  National Institute of Justice Digital Evidence Policies and Procedures Manual~\cite{nij}
        \item[NIST] NIST Internal Report 8354 Digital Investigation Techniques~\cite{nist_ir_8354}
    \end{tablenotes}
    \end{threeparttable}

\end{table}

The first step in SPADA involves defining the values for the five variables mentioned in Section~\ref{sec:spada}. 

Table~\ref{fig:variables} lists the sources of documentation (S) selected for the construction of the threat model. 
These include some of the works that focus on privacy threats and/or risks in DFCI, as presented in Section~\ref{sec:related-work}.
We extend such list with relevant standards and best-practices in the field of digital forensics and crime investigation.

The main property (P) that we target is privacy, in all its facets, and hence we consider both hard and soft privacy. 
While hard privacy focuses on minimising the risks associated with the collection and retention of personal data, soft privacy focuses on the appropriate use and sharing of personal data while respecting individuals' rights to control their data.

The application domain (A) is, naturally, that of digital forensics in crime investigation (DFCI). 

Furthermore, we desire a consistent, abstract level of detail (D), so as to focus on the broader picture.

As for the threat agents (A), we consider all potential agents who could raise privacy threats in digital forensics, including attackers, data controllers/processors, and third parties. 
In the context of DFCI, these roles could be occupied by various individuals or entities involved in the investigation, such as investigators, judges, or external parties like cloud service providers. 
It is important to note that these roles are often fluid, as an investigator may impersonate a data controller, or a third party may assume the role of an attacker if certain safeguards are not in place.

\subsection{Step 1:~Domain-Independent Threat Elicitation}

\begin{figure}[ht]
    \centering
\includegraphics[width=0.6\textwidth]{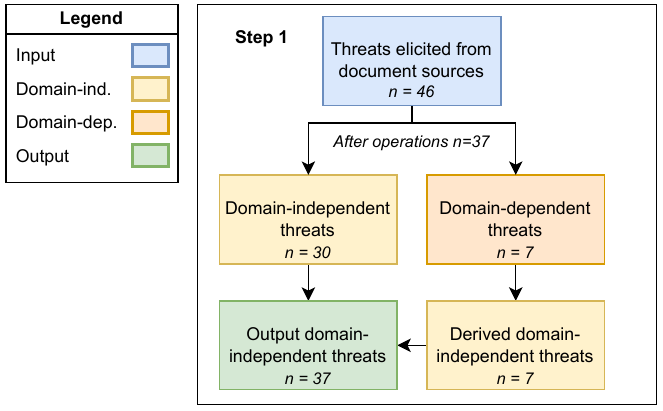}
    \caption{PRISMA-like visualisation of Step 1.}
    \label{fig:step1}
\end{figure}

The next step in SPADA involves eliciting domain-independent threats. Figure~\ref{fig:step1} illustrates this step in a PRISMA-like diagram.
We collect 46 threats from the listed document sources and then apply the SPADA operations to refine the list as follows.
We discard a total of three threats, as they fall out of the scope of this study: the threats ``Malicious code injection'', ``Denial of service attack'', and ``Loss of encryption key'' refer to a security aspects rather than privacy, which is our target property. 
In addition, since some of the threats can be arguably deemed redundant, we employ the TEAM 3 algorithm~\footnote{\url{https://github.com/tsumarios/TEAM}}, which applies the embrace operation in an automated fashion. The details of the algorithm are part of a study that is yet to be published.
For example, ``Unauthorized person access to the big data forensic platform'' and ``No systematic monitoring of authorizations'' can be embraced into a single threat ``Lack of authorization management''.
As a result, we further reduce the list to 37 threats.

\subsubsection{Domain-dependency handling}
Moreover, we observe the presence of threats that are already domain-dependent in the document sources. This is not suprising as the selected sources are were directly related to specific elements within the digital forensic process in the context of criminal investigations.
While some of the threats may be transformed into a domain-independent counterpart within the embrace operations (see the previous example where ``Lack of authorization management'' results from the application of the embrace operation to a domain-dependent threats), it may also happen that other threats remain specific to the application domain. For example, the result of an embrace with the threat ``Investigation report (paper documents) sent to wrong destination'' produced ``Misdirection of investigation documents'', which still maintains the domain specificity.
For this reason, we can directly retain the (7 out of 37) domain-dependent threats in the output list, which will be extended in Step 3, as we shall detail below.

Notably, the fact that document sources already present domain-dependent threats allows for the application of SPADA in a new, unexplored direction. 
In fact, we can apply the rename operation to remove the domain-dependency from these threats, keeping in mind that a domain-independent threat is a threat that is universally applicable to any domain.
For example, the threat ``Lack of support for privacy management by forensic tool vendors'' can be renamed into ``Lack of support for privacy management by software vendors''. 
By following this approach, we can derive a subset of domain-independent threats that can extend the list of domain-independent threats to consider in Step 3. 
This derivation process can be done both before refining the input list, with the original, collected threats, and also after refining the input list. 
Naturally, as there may be redundancy among the derived domain-independent threats in the pre-refinement and post-refinement lists, it is appropriate to apply the SPADA operations to merge the derived domain-independent lists into one.
For the sake of brevity, Table~\ref{tab:domain-dep-threats} illustrates the (12 out of 46) pre-identified domain-dependent threats and the domain-independent threats derived from the rename operations as described above, before the refinement -- the full results online~\cite{repo} also include the derived threats (7 out of 37) after the refinement.
At the end of Step 1, we obtain a total of 30 domain-independent threats.

\begin{table*}[ht]
\caption{Derivation of domain-independent privacy threats from domain-dependent ones before refining the input list.}
\label{tab:domain-dep-threats}
\centering
\begin{tabular}{p{3cm} p{12cm}}
\toprule
\textbf{Source of doc.} & \textbf{Threat (Domain-Dependent) → Threat (Domain-Independent)} \\
\midrule
\multirow{6}{*}{Seyyar et al.} 
  & Data process/read for wrong case → Improper data processing or access \\
  & Unauthorized person access to the big data forensic platform → Unauthorized person access to the big data platform \\
  & Investigation report (paper documents) sent to wrong destination → Misdelivery of confidential document \\
  & Access to data after case is closed → Access to data beyond retention period \\
  & Authorizations not granted at case level → Insufficient access control mechanisms \\
  & Errors while uploading seized digital material → Errors in data upload or ingestion \\
\midrule
Chaure et al.
  & Erroneous allegations due to deleted files → Erroneous allegations due to deleted files \\
\midrule
\multirow{5}{*}{Rowe} 
  & Unwarranted reporting of forensic findings → Unwarranted reporting of findings \\
  & Surreptitious searches → Covert or unlawful data searches \\
  & Selling of private forensic data → Illicit sale of private data \\
  & Criminal use of digital forensics → Malicious misuse of practice \\
  & Lack of support for privacy management by forensic tool vendors → Lack of support for privacy management by software vendors \\
\bottomrule
\end{tabular}
\end{table*}

\subsection{Step 2:~Domain-Dependent Asset Collection}

\begin{figure}[ht!]
    \centering
\includegraphics[width=0.5\textwidth]{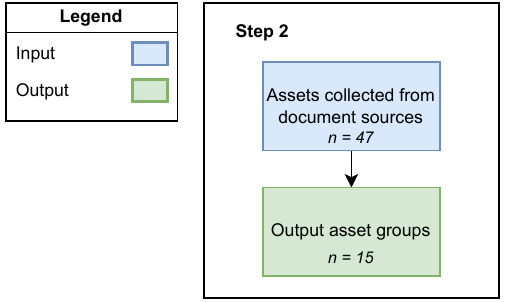}
    \caption{PRISMA-like visualisation of Step 2.}
    \label{fig:step2}
\end{figure}

We proceed to Step 2, where domain-dependent assets are collected. Figure~\ref{fig:step2} illustrates this step in a PRISMA-like diagram.
The document sources that we consider do not explicitly provide a taxonomy or list of assets, hence we build such a list upon our scrutiny. 
Specifically, we analyse the document sources and identify the candidate assets. 
Then, we apply the SPADA operations, in particular the embrace, to gather similar assets under groups. 
This also aligns with the desired level of detail:~asset groups result more abstract than single items. 

We establish a total of 15 asset groups from the document sources. 
Table~\ref{tab:assets} illustrates the collected assets. 
For example, we consider physical storage devices, hard drives, papers and documents under the ``Storage media'' group. 
``Email and messaging'' includes digital communications (e.g., WhatsApp, Signal, etc.), whereas ``Forensic tools and equipment'' gathers both forensic software (e.g., FTK, EnCase, Autopsy, etc.) and hardware (e.g., write blockers, etc.). 

\begin{table}[ht]
\caption{Assets collected in Step 2.}
\label{tab:assets}
\centering
\begin{tabular}{ll}
\toprule
\textbf{Source of Documentation} & \textbf{Asset} \\
\midrule
ISO & Storage media \\
CoE DF, IPOL & Cloud and remote storage \\
CoE DF & Email and messaging \\
CoE DF & Communication and network logs \\
CoE DF & Authentication and access logs \\
CoE DF, NIST & Forensic tools and equipment \\
CoE DF, NIJ & Case management databases \\
CoE DF & Secure forensic workstations \\
CoE DF & Forensic lab \\
CoE EEG & Desktop devices \\
CoE EEG & Mobile devices \\
CoE EEG & IoT devices \\
CoE EEG & Location and tracking data \\
CoE EEG & Cryptocurrency data \\
IPOL & System and application logs \\
\bottomrule
\end{tabular}
\end{table}

\subsection{Step 3:~Domain-Dependent Threat Elicitation}

\begin{figure}[ht]
    \centering
\includegraphics[width=0.6\textwidth]{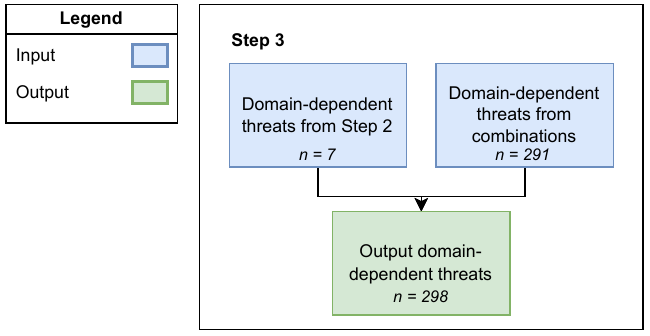}
    \caption{PRISMA-like visualisation of Step 3.}
    \label{fig:step3}
\end{figure}

We complete the application of SPADA with the instantiation of the domain-independent threats from Step 1 into domain-specific threats, by combinations with the assets from Step 2. Figure~\ref{fig:step3} illustrates this step in a PRISMA-like diagram.

For example, the threat ``Unauthorized access to data'' may affect all the assets, hence we can instantiate a total of 15 domain-dependent threats by combining it with all the assets (which are indeed 15). We elicit 291 domain-dependent threats by combinations. Such list can be enriched by adding the pre-identified domain-dependent threats saved in Step 1.
As a result, we obtain a total of 298 domain-dependent threats.

\section{A Privacy Threat Model for DFCI}
\label{sec:threat-model}

This section presents the resulted privacy threat model for digital forensics in crime investigation. For the sake of brevity, Table~\ref{tab:threat-model} features an extract of the privacy threat model.

\begin{table*}[ht]
\caption{Extract of the privacy threat model for DFCI.}
\label{tab:threat-model}
\centering
\begin{tabular}{p{6cm} p{4.5cm} p{4.5cm}}
\toprule
\textbf{Threat (Domain-Independent)} & \textbf{Asset(s)} & \textbf{Threat Agent(s)} \\
\midrule
Poor training & 
All assets & 
Data Controller, Third Party \\
\midrule
Cross-border data privacy concerns & 
Cloud and remote storage, \newline 
Email and messaging, \newline
Case management databases, \newline
Location and tracking data & 
Data Controller, Data Processor, Third Party \\
\midrule
Lack of privacy management & 
Forensic tools and equipment, \newline
Secure forensic workstations, \newline
Case management databases & 
Data Controller, Data Processor, Third Party \\
\midrule
\textbf{Threat (Domain-Dependent)} & \textbf{} & \textbf{Threat Agent(s)} \\
\midrule
Errors while uploading seized digital material & 
 & 
Data Processor, Third Party \\
\midrule
Selling of private forensic data & 
 & 
Attacker, Data Controller, Data Processor, Third Party \\
\bottomrule
\end{tabular}
\end{table*}

For example, the threat ''Cross-border data privacy concerns'' affecting the assets ``Cloud and remote storage, Email and messaging, Case management databases, Location and tracking data'', may be raised by data controller, data processor and/or third party. A data controller/processor (e.g., law enforcement agency) may unknowingly violate privacy laws by storing evidence in a jurisdiction with weaker protections. Similarly, a third-party cloud provider might not enforce strong access controls, allowing attackers or unauthorized entities to access forensic data.
Furthermore, the threat ``Selling of private forensic data'' may be raised by each of the threat agents considered in the previous section. Forensic data often contains highly sensitive personal information, including communications, financial transactions, and digital activities of suspects, victims, and unrelated individuals. A data processor (e.g., a forensic expert with privileged access to case data) may be tempted to sell such data to malicious actors, such as criminal groups or investigative journalists. Similarly, attackers could infiltrate forensic systems to exfiltrate and monetise these data on dark web marketplaces.

Moreover, many threats in the model stem from procedural and governance gaps rather than purely technical vulnerabilities. For example, the threats ``Errors while uploading seized digital material'' or ``Lack of privacy management'' indicate that improper handling of forensic data by data controller/processor or third party may result in privacy violations. These systemic risks suggest that improving operational procedures and compliance frameworks is just as critical as implementing technical safeguards.

\section{Discussion}
\label{sec:discussion}

Privacy threat modelling can help forensic investigators identify potential violations early in the process, thus enabling them to implement safeguards that minimise risk and, consequently, maintain forensic practices ethically sound and legally compliant.
As a result of the application of SPADA in DFCI, we constructed a privacy threat model based on two lists of privacy threats. 
The first is a list of domain-dependent privacy threats for DFCI. 
While we cannot claim that our list is definitive and complete, it significantly expands upon previous works.
The second list composes a general knowledge base that can expand existing taxonomies in support of the pursuit to completeness and application to any domain. 
To this extent, the second list can be leveraged to extend the already existing privacy threat knowledge base offered by SPADA~\cite{spada_repo}.

\paragraph{Partial Validation}
The authoritative nature of the document sources that we leveraged in the construction of the resulting threat model can support a partial validation of the results. 
Such partial validation can be further extended and confirmed by searching for news, by means of web queries, thereby checking the tangibility of the proposed threats in real-world scenarios. 
For example, the threat ``Surreptitious searches'' matches with the news of Serbian authorities that were using forensics tools to surveil activists and journalists, as presented in Section~\ref{sec:introduction}.

\paragraph{Limitations}
As part of SPADA, some steps relied on subjective judgment (e.g., embracing arguably redundant threats). This was partially mitigated by adopting the TEAM 3 algorithm.
We argue that applying the proposed privacy threat model to real-world forensic investigations can further assess its effectiveness in guiding privacy-conscious forensic practices.
However, one of the main challenges in such validation process is the variability of privacy laws across jurisdictions. 
In fact, different countries have different regulations regarding data access, retention, and consent, which makes it difficult to create a universal privacy threat model.
Moreover, real-world constraints, e.g., time pressure and resource scarcity, in forensic operations may limit the feasibility of implementing certain privacy controls.
For this reason, the threats elicited in Section~\ref{sec:application} may require further analysis and refinement to be properly instantiated, with a weight factor that indicates the relevancy of the threat in reference to the legal and regulatory context of each jurisdiction.

\section{Conclusions}
\label{sec:conclusions}

The increasing reliance on digital forensics in crime investigations (DFCI) may pose significant privacy concerns. 
This paper addressed the challenge of modelling privacy threats in this domain. The research question found an answer in the application of the SPADA methodology, which produced a privacy threat model for DFCI. The model was built from a two-fold approach that yielded a list of domain-dependent privacy threats for the DFCI domain and a list of domain-independent privacy threats. The application of SPADA to this domain demonstrated how SPADA assists in handling domain-dependency during threat elicitation.

Our future work looks at: ranking threats by impact and likelihood to support prioritisation within the proposed threat model;  
further reduce subjectivity in SPADA through automation (e.g., combining the TEAM algorithms with Large Language Models); and continue the formalisation of DFCI, expanding on anti-digital forensics and cybersecurity threats.

This paper ultimately encourages interdisciplinary collaboration between forensic practitioners, legal experts, and privacy researchers to further strengthen the alignment between digital forensic practices and privacy-preserving principles.

\subsection*{Acknowledgements}

G. Bella acknowledges: IT PNRR 2022 SERICS Spoke 7, BaC, Project ``SCAR4SUD - SCAR’s Four Security-Unravelling Dimensions'', CUP C69J24000320008.
M. Raciti acknowledges: IT PNRR 2022 SERICS Spoke 6, Task 1.2, Project ``SCAI — Supply Chain Attack Avoidance''.

\bibliographystyle{unsrtnat}
\bibliography{references}  






\end{document}